\documentclass[a4paper,11pt]{article}

\usepackage{jcappub} 

\usepackage[T1]{fontenc} 
\usepackage{natbib}
\usepackage{amssymb}
\usepackage{amsmath}
\usepackage{amsfonts}
\usepackage{graphicx}
\usepackage[hang,tight,raggedright]{subfigure}
\usepackage{multirow}
\usepackage{subfigure}
\usepackage{float}
\usepackage{graphics}
\usepackage{slashed}
\usepackage{color}
\usepackage{bm}
\usepackage{braket}
\usepackage{psfrag}
\usepackage{booktabs}
\usepackage{latexsym}
\usepackage{pythonhighlight}

\usepackage{setspace}
\usepackage{amsmath}
\usepackage{amsthm}
\usepackage{mathrsfs}
\usepackage{epsf}
\usepackage{epsfig}
\usepackage{bbm}
\usepackage{bm}
\usepackage{amsfonts}
\usepackage{amssymb}
\usepackage{latexsym}
\usepackage[english]{babel}
\usepackage{multirow}
\usepackage{float}
\usepackage{url}
\usepackage{slashed}
\usepackage{xcolor} 
\usepackage[utf8]{inputenc}
\usepackage{stmaryrd} 
\usepackage{enumitem}
\usepackage{siunitx}
\usepackage{verbatim}
\usepackage{bbm}

\usepackage{caption3}
\usepackage{appendix}

\makeatletter

\newcommand{\Rmnum}[1]{\expandafter\@slowromancap\romannumeral #1@}
\makeatother

\newcommand{\be}{\begin{equation}}
	\newcommand{\ee}{\end{equation}}
\newcommand{\ba}{\begin{array}}
	\newcommand{\ea}{\end{array}}
\newcommand{\bea}{\begin{eqnarray}}
	\newcommand{\eea}{\end{eqnarray}}
\newcommand{\nn}{\nonumber}

\usepackage{theorem}
{
	\theoremheaderfont{\bfseries}
	\theorembodyfont{\normalfont}
}

\title{Implication of nano-Hertz stochastic gravitational wave background on ultralight axion particles}
\author[a]{Jing Yang,}
\author[a]{Ning Xie,}
\author[a,1]{and Fa Peng Huang\note{Corresponding author.}}
\emailAdd{huangfp8@sysu.edu.cn}
	
\affiliation[a]{MOE Key Laboratory of TianQin Mission, TianQin Research Center for Gravitational Physics \& School of Physics and Astronomy, Frontiers Science Center for TianQin, Gravitational Wave Research Center of CNSA, Sun Yat-sen University (Zhuhai Campus), Zhuhai 519082, China}

\abstract{Recently, the Hellings Downs correlation has been observed by different pulsar timing array (PTA) collaborations, such as NANOGrav, European PTA, Parkes PTA, and Chinese PTA.
These PTA measurements of the most precise pulsars within the Milky Way show the first evidence for the stochastic gravitational wave background of our Universe.
We study the ultralight axion interpretation of the new discovery by investigating the gravitational wave from axion transitions
between different energy levels of the gravitational atoms, which are composed of cosmic populated Kerr black holes and their surrounding axion clouds formed through the superradiant process.
By Bayesian analysis,
we demonstrate that 
this new observation naturally admits an ultralight axion interpretation around $10^{-21}$~eV, which 
is consistent in magnitude with the typical mass of fuzzy dark matter.
}


\begin{document}
\maketitle
\flushbottom

\section{Introduction}

Recently, the expected Hellings-Downs~\cite{Hellings:1983fr} curve has finally been observed by the global pulsar timing array (PTA) experiments including NANOGrav~\cite{NANOGrav:2023gor}, European PTA (EPTA)~\cite{EPTA:2023fyk},
Parkes PTA (PPTA)~\cite{Reardon:2023gzh}, and Chinese PTA (CPTA)~\cite{Xu:2023wog}, which provide strong evidence for the detection of stochastic 
gravitational wave background (SGWB).
The observation of SGWB has wide and deep implications for astrophysics, particle physics, and cosmology. Especially, the PTA measurement of SGWB opens new windows to explore the new physics beyond the standard model~\cite{NANOGrav:2023hvm, EPTA:2023xxk}.
In this work, we focus on the implication of the new observation to the ultralight
axion~\cite{Peccei:1977hh,Weinberg:1977ma,Wilczek:1977pj,Kim:1979if,Shifman:1979if,Zhitnitsky:1980tq,Dine:1981rt,Sikivie:2020zpn} or axion-like particles~\cite{Svrcek:2006yi} (in this letter, we simply use axion to represent both axion and axion-like particle), which could provide a natural solution to the strong CP problem and a well-motivated dark matter (DM) candidate~\cite{Preskill:1982cy, Sikivie:2006ni,Abbott:1982af,Dine:1982ah}. 
Motivated by the current situation of DM search, recently more and more studies have focused on ultralight bosons~\cite{Du:2022trq}, such as ultralight axions. 
As pseudo scalar particles, ultralight axion particles can form Bose-Einstein condensation and further
form dense axion objects, like 
axion stars or axion clusters. Especially, the so-called axion cloud can be formed through the superradiant process around Kerr black hole (BH)~\cite{Penrose:1969pc, zeld, Starobinsky:1973aij, Detweiler:1980uk}. 
Once the axion cloud is formed, abundant signals could be produced.
For example,
the axion cloud can be a promising gravitational wave (GW) source~\cite{Brito:2015oca,Arvanitaki:2014wva,Arvanitaki:2016qwi,Dev:2016hxv,Hannuksela:2018izj,Baryakhtar:2017ngi,Xie:2022uvp}, which can be used to explore the properties of axion particles~\cite{Brito:2017zvb, Arvanitaki:2010sy, Brito:2014wla, Yoshino:2013ofa, Yang:2023vwm}. 

In this article, we fit the first observation of the nano-Hertz SGWB with the ultralight axion particles by considering the transition process of the bound state of the axion cloud and Kerr BH, which is ignored in the previous study.
We introduce the gravitational atom (GA) as the new type of interpretation of nano-Hertz SGWB and calculate the GW radiation power from the isolated GA. Further, considering the cosmic BH population, we give the expected SGWB for the ultralight axion in the mass range of fuzzy DM~\cite{Hu:2000ke,Hui:2016ltb,Drlica-Wagner:2022lbd}, and compare it with the new PTA data and SGWB from super-massive black holes binaries (SMBHBs) by Bayesian analysis.
%

\section{The axion cloud and gravitational atom from superradiance around Kerr black hole}

From general relativity, the BH superradiant process~\cite{Penrose:1969pc, zeld, Starobinsky:1973aij, Detweiler:1980uk} can naturally form the axion cloud around the Kerr BH since there exists an ergosphere between the static limit surface and the event horizon. From quantum field theory, the axion field is described by the Klein-Gordon equation in the Kerr metric. After solving this equation, one can obtain the frequency eigenvalues. It is shown that the frequency of the axion field has a positive imaginary part, which leads to an exponential growth solution when the Compton wavelength of axion particle is comparable with the BH size~\cite{Detweiler:1980uk,Dolan:2007mj}. 
 And if their initial spin is greater than the critical value that triggers the superradiant process, the axion cloud extracts energy and angular momentum from the Kerr BH.
The GA, similar to the hydrogen atom, consisting of the condensed axion cloud and BH 
can be generated. And axions can occupy different energy levels of the GA to form bound states.
Once the GA is formed, various types of GW signals~\cite{Brito:2015oca,Arvanitaki:2014wva,Arvanitaki:2016qwi,Hannuksela:2018izj,Baryakhtar:2017ngi,Dev:2016hxv,Xie:2022uvp} can be produced from axions annihilating to gravitons or transitions between different energy levels of the GA.
\begin{figure*}[ht]
	\begin{center}
		\includegraphics[width=0.8\linewidth,clip]{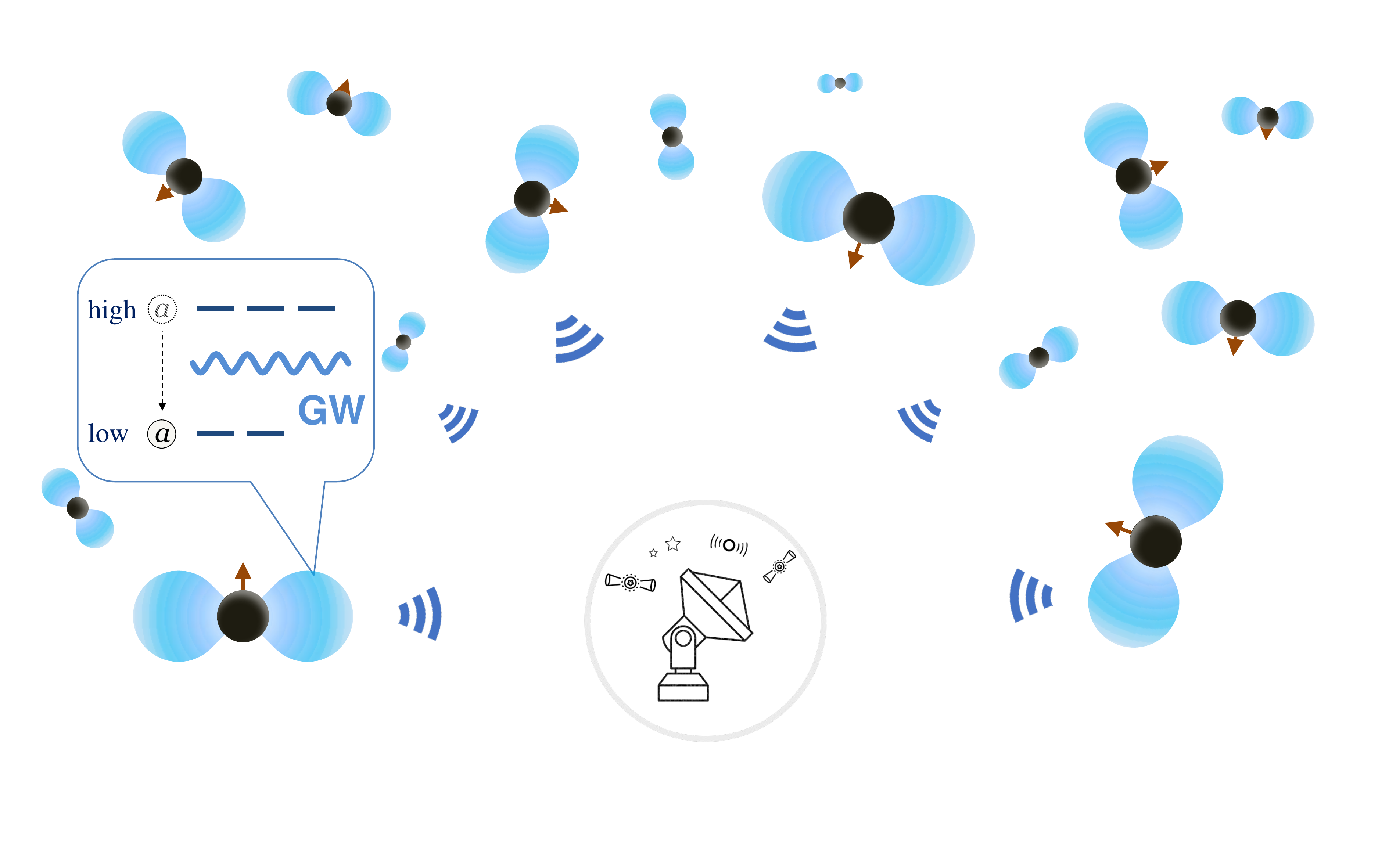}\\
		\caption                    {The cosmic populated SMBHs dressed with the  axion cloud as a natural source of nano-Hertz SGWB. The energy-level transition process can radiate GWs continuously, which naturally fall in the  nano-Hertz frequency band. Consequently, the PTA could  detect  this new source which provides a new approach to probe ultralight axion DM and isolated BHs.  
                    }                
		\label{sc}
	\end{center}
\end{figure*}
If the occupation numbers of the energy levels are sufficiently large, the GW effect could be detectable. 
In our Universe, there distributes a large number of super-massive black holes (SMBHs).
The axion transitions around the SMBHs could contribute to the signal of nano-Hertz SGWB, which is detectable through PTA experiments, as shown in Fig.~\ref{sc}. 
It is worth noticing that the SGWB from energy-level
transition is not monochromatic due to the fact that the astrophysical SMBHs could have a wide mass range.
Meanwhile, the SGWB observation might help to explore the ultralight axion DM. 


\section{The radiation power of an isolated gravitational atom}

We denote the BH, axion mass as $M_b$ and $m_a$ respectively. 
The fine structure constant of the GA is $\alpha=GM_bm_a$ where $G$ is Newton's constant. The gravitational radius of BH is $r_g=GM_b$. We use $|nlm\rangle$ to represent energy levels where $n$,~$l$,~$m$ are the principal, orbital, and magnetic quantum numbers.  

\begin{figure}[!htp]
	\begin{center}
		\includegraphics[width=0.7\linewidth]{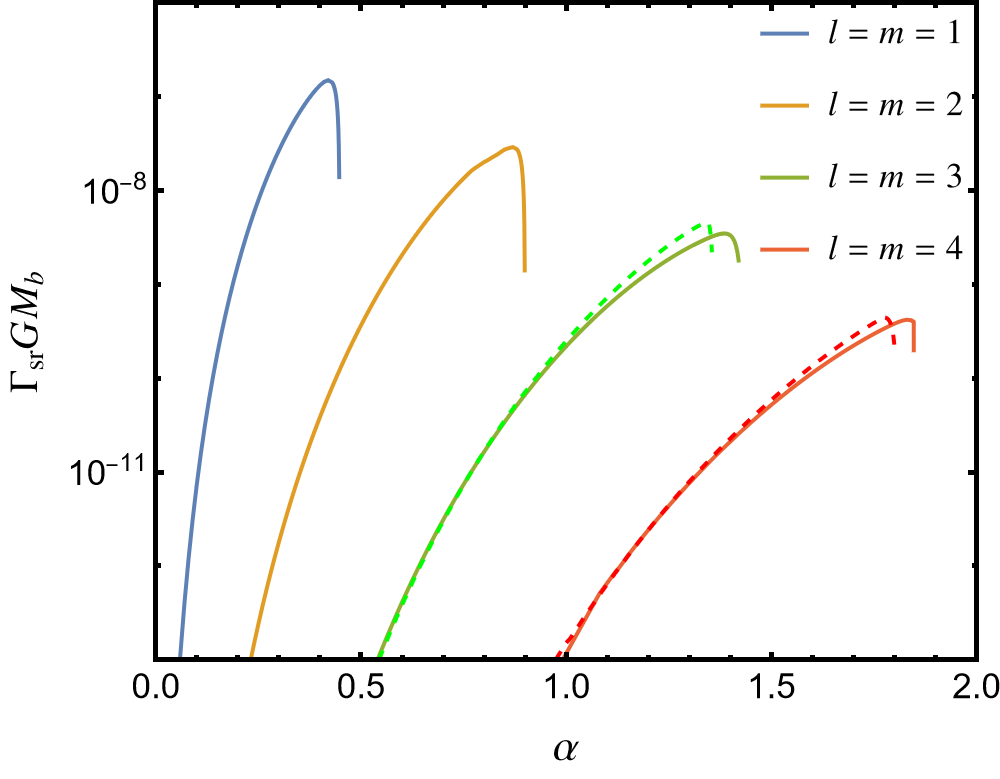}
		\caption{The superradiant rate of different energy levels in the unit $1/GM_b$ with the BH spin $J=0.99~GM_b^2$. The thick lines are results calculated with $l=m=n-1$ and the dashed lines are the results of the energy levels $|533\rangle$ and $|644\rangle$. }\label{fig:sr}
	\end{center}
\end{figure}

The superradiant condition~\cite{Arvanitaki:2014wva} for a level with frequency $\omega$ is,
\be
\frac{\omega}{m}<\frac12\left(\frac{a}{1+\sqrt{1-a^2}} \right)r_g^{-1}
\ee
where $a=J/(GM_b^2)$ and $J$ is the spin angular momentum of the BH.
When the superradiant condition is satisfied, the occupation number grows as 
\be
\frac{dN}{dt}\Big|_{\text{sr}}=\Gamma_{\text{sr}}N,
\ee
where $\Gamma_{\text{sr}}$ is the superradiant rate.
The superradiant rate of different energy levels is calculated numerically following the method used in Refs.~\cite{Dolan:2007mj,Yoshino:2013ofa}. The results of energy levels $l=m=n-1$ as well as $|nlm\rangle=|533\rangle,|644\rangle$ when $J=0.99~GM_b^2$  are shown in Fig.~\ref{fig:sr}. The superradiant time scale is
\be
\tau_{\text{sr}}=1/\Gamma_{\text{sr}}=1.57\times 10^5~\text{yr}~\frac{10^{-10}}{\Gamma_{\text{sr}}GM_b}\left(\frac{M_b}{10^8~M_{\odot}}\right).
\ee
To get a stronger SGWB signal, more axions are needed to occupy the initial state energy level for the transition process. The energy levels of the transition process $|533\rangle\rightarrow |433\rangle$ have the smallest principle and orbital quantum numbers when the superradiant rate of the initial state energy level is larger than the final state energy level. We consider the dominant transition processes $|533\rangle\rightarrow |433\rangle$ and $|644\rangle\rightarrow |544\rangle$ in this work since the superradiant rate for the transition process with higher energy levels is slower compared with the $|533\rangle\rightarrow |433\rangle$ and $|644\rangle\rightarrow |544\rangle$ processes~\cite{Arvanitaki:2014wva}. 

We can see from Fig.~\ref{fig:sr}, the superradiant process is dominated by energy levels $|533\rangle$ and $|644\rangle$ at  $0.90 \lesssim\alpha\lesssim 1.34$ and $1.38 \lesssim\alpha\lesssim 1.79$, respectively. We use $\Gamma_{\text{sr,i}}$ and $\Gamma_{\text{sr,f}}$ to represent the superradiant rate of the initial and final state energy levels for the transition process respectively. The relevant scales are the accretion time scale $t_S$, the superradiant time scale $\tau_{\text{sr}}=1/\Gamma_{\text{sr,i}}$ and the age of the universe $t_0$. The time of the superradiant process must be smaller than $t_S$ and $t_0$. If the superradiant rate is fast enough, the superradiant situation is saturated with the total axion number 
\be
N_{\text{max}}\simeq \frac{GM_b^2}{m}\Delta a,\label{Nm}
\ee
where $\Delta a=\Delta J/GM_b^2\simeq 0.04$ is the difference between the initial and final BH spin with the initial spin $J=0.99~GM_b^2$. We consider the axion number of the initial state energy levels and final state energy levels as $N_i\simeq\text{min}[N_{\text{max}},\text{exp}(\text{min}[t_0,t_S]*\Gamma_{\text{sr,i}})]$ and $N_f=N_i^{\Gamma_{\text{sr,f}}/\Gamma_{\text{sr,i}}}$ respectively when the superradiant process is terminated.
 
The transition rate is given by~\cite{Arvanitaki:2014wva},
\begin{equation} \label{quad}
\frac{dN_i}{dt}\Big |_{\text{tr}}=-A_{nml,n'm'l'}N_iN_f\frac{G\alpha^9}{r_g^3},
\end{equation}
where $A_{644,544}\approx3\times 10^{-7}$ and $A_{533,433}\approx1.2\times 10^{-6}$ for the transition process $|644\rangle\rightarrow |544\rangle$ and $|533\rangle\rightarrow |433\rangle$, respectively.
If the axion annihilation process is considered after superradiance is terminated, the axion numbers evolve as
\begin{align}
\frac{dN_i}{dt}&=-\Gamma_{\text{tra}}N_i N_f-\Gamma_{\text{ann},i}N_i^2, ~\\
\frac{dN_f}{dt}&=\Gamma_{\text{tra}}N_i N_f-\Gamma_{\text{ann},f}N_f^2,
\end{align}
where the transition rate $\Gamma_{\text{tra}}=A_{nml,n'm'l'}~G\alpha^9/r_g^3$ can be read from Eq.~(\ref{quad}) and $\Gamma_{\text{ann},i} 
  (\Gamma_{\text{ann},f})$ is the annihilation rate of the initial (final) state energy level. The $\Gamma_{\text{ann}}$ can be estimated as~\cite{Yoshino:2013ofa}
\begin{equation}
\Gamma_{\text{ann}}= C_{nl}\frac{G\alpha^{p}}{r_g^3},
\end{equation}
where $p=4l+11$, and
\be
C_{nl}=\frac{16^{l+1}l(2l-1)\Gamma(2l-1)^2\Gamma(l+n+1)^2}{n^{4l+8}(l+1)\Gamma(l+1)^4\Gamma(4l+3)\Gamma(n-l)^2}.
\ee

Since $\Gamma_{\text{ann},f}N_f\ll\Gamma_{\text{tra}}N_i$, we can neglect the annihilation process at the final state energy level. For the evolution of the initial state energy level, we need to consider different cases: \\
\begin{itemize}
    \item \textbf{Case 1:} $\Gamma_{\text{ann},i}N_i(t_{\star})>\Gamma_{\text{tra}}N_f(t_{\star})$\\
    \hspace*{1em} Here $t_{\star}$ is the time when superradiance is terminated. The annihilation process dominated the evolution of the axion number in the first stage, 
\bea
\frac{dN_i}{dt}&\simeq &-\Gamma_{\text{ann},i}N_i^2,~\\
\frac{dN_f}{dt}&=&\Gamma_{\text{tra}}N_i N_f,
\eea
the solution is
\bea
N_i(t_{\star}+t)&=&\frac{N_i(t_{\star})}{1+\Gamma_{\text{ann},i}N_i(t_{\star})t},~\\
N_f(t_{\star}+t)&=&N_f(t_{\star})(1+\Gamma_{\text{ann},i}N_i(t_{\star})t)^{\Gamma_{\text{tra}}/\Gamma_{\text{ann},i}}.
\eea
Then we can solve the equation $\Gamma_{\text{ann},i}N_i(t_{\star}+t)=\Gamma_{\text{tra}}N_f(t_{\star}+t)$ to get the time length $t_a$ of the first stage, we obtain
\bea
t_a&=&(N_i(t_{\star})\Gamma_{\text{ann},i})^{-\frac{\Gamma_{\text{tra}}}{\Gamma_{\text{tra}}+\Gamma_{\text{ann},i}}}(N_f(t_{\star})\Gamma_{\text{tra}})^{-\frac{\Gamma_{\text{ann},i}}{\Gamma_{\text{tra}}+\Gamma_{\text{ann},i}}} \nn \\
&-&\frac{1}{N_i(t_{\star})\Gamma_{\text{ann},i}}.
\eea
Then the second stage is that the transition process dominates the evolution of axion number. Since $N_i(t_{\star}+t_a)\gg N_f(t_{\star}+t_a)$, we can roughly take $N_i$ unchanged in this stage. Thus the occupation number of the final state energy level evolves as $N_f(t_{\star}+t_a+t)=N_f(t_{\star}+t_a)e^{\Gamma_{\text{tra}}N_i(t_{\star}+t_a)t}$. The time length of the second stage is estimated as $t_b=\ln\big(N_i(t_{\star}+t_a)/N_f(t_{\star}+t_a)\big)/\big(\Gamma_{\text{tra}}N_i(t_{\star}+t_a)\big)$. The gravitational radiation time scale is $\tau_{\text{GW}}=t_a+t_b$.
\end{itemize}

\begin{itemize}
    \item \textbf{Case 2:} $\Gamma_{\text{ann},i}N_i(t_{\star})<\Gamma_{\text{tra}}N_f(t_{\star})$\\
\hspace*{1em} In this case, the transition process would dominate the evolution. The axion number of the final state energy level is $N_f(t_{\star}+t)=N_f(t_{\star})e^{\Gamma_{\text{tra}}N_i(t_{\star})t}$. The gravitational radiation time scale is estimated as $\tau_{\text{GW}}=\ln(N_i(t_{\star})/N_f(t_{\star}))/(\Gamma_{\text{tra}}N_i(t_{\star}))$.
\end{itemize}

The energy spectrum of the GA is given by
\begin{equation}
 \omega_n\approx m_a(1-\frac{\alpha^2}{2n^2}). 
\end{equation}
We can calculate the radiation power of the transition process as
\begin{equation}
P=-\frac{dE}{dt}=\frac{dN_f(t)}{dt}\Delta \omega,  
\end{equation}
where $\Delta \omega$ is the frequency split of the initial and final states.
Thus the radiation energy of GW from axion transitions between different energy levels  is
\begin{equation}\label{E}
\begin{aligned}
 E_{t}(t)&=(N_f(t)
 -N_f(t_{\star}))\Delta \omega.        
\end{aligned}
\end{equation}

\section{Stochastic gravitational wave with cosmic black hole population}\label{sec3}
The distribution of SMBHs is essential to calculate the spectra of the SGWB. 
For isolated BHs, the number density in the comoving volume is estimated as~\cite{Babak:2017tow,Gair:2017ynp}
 \begin{equation}\label{distr}
 \frac{dn}{d \log_{10}M_b}=0.002\left(\frac{M_b}{3\times 10^6~ M_{\odot}}\right)^{0.3}\text{Mpc}^{-3},
 \end{equation}
 where 
 $M_{\odot}$ is the solar mass. This formula is valid for $10^4~M_{\odot}<M_b<10^7~M_{\odot}$, while for $M_b>10^{7}~M_{\odot}$, we reduce the density 100 times for a conservative model compared to Eq.~(\ref{distr}). As for the SMBH spin model, a large spin is required to produce sufficient axions through the superradiant process, and the simulations of thin disk accretion find that $70\%$ of SMBHs have extreme initial spin~\cite{Volonteri:2004cf,Sesana:2014bea}. Thus, we assume $70\%$ of SMBHs with initial spin $a=J/GM_b^2=0.99$, and we neglect GW signals from other SMBHs with smaller spin.
The accretion time is estimated via Salpeter time~\cite{Brito:2017zvb},
\be
t_S=4.5\times 10^8~\text{yr}~\frac{\eta}{f_{\text{Edd}}(1-\eta)}.
\ee
where we take the Eddington ratio $f_{\text{Edd}}=0.01$ in this work and $\eta$ is the thin-disk radiative efficiency factor~\cite{Bardeen:1972fi},
\bea
\eta&=&1-\sqrt{1-\frac{2}{3r}}, \\
r&=&3+Z_2-\sqrt{(3-Z_1)(3+Z_1+2Z_2)}, \\
Z_1&=&1+(1-a^2)^{1/3}[(1+a)^{1/3}+(1-a)^{1/3}], \\
Z_2&=&\sqrt{3a^2+Z_1^2}.
\eea

\begin{figure}[!htp]
	\begin{center}
			\includegraphics[width=1\linewidth]{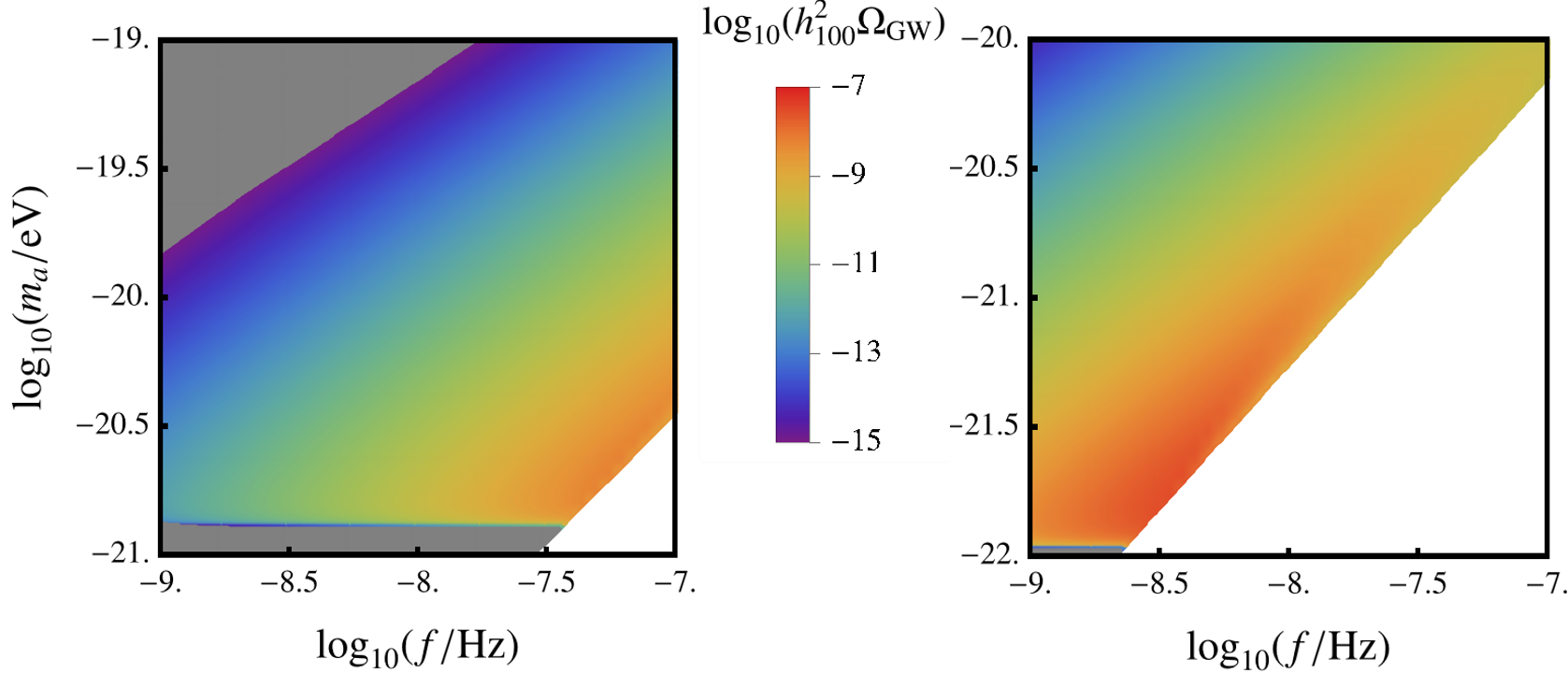}
			\caption{Left panel: the SGWB energy density from axion transition process $|644\rangle\rightarrow |544\rangle$ with the axion mass range $10^{-21}\sim 10^{-19}~\text{eV}$. Right panel: the SGWB energy density from axion transition process $|533\rangle\rightarrow |433\rangle$ with the axion mass range $10^{-22}\sim 10^{-20}~\text{eV}$. The gray region represents the GW strength $h_{100}^2\Omega_{\text{GW}}<10^{-15}$ while the blank region depicts no GW signal in the corresponding parameter space. }\label{fig:final}
		\end{center}
\end{figure}

The SGWB from populated 
GAs is derived as~\cite{Phinney:2001di}
\begin{equation}
	\Omega_{\mathrm{gw}}(f)=0.7\times\frac{f}{\rho_c} \int d M_b d z \frac{d t}{d z} \frac{d \dot{n}}{d M_b } \frac{d E_s}{d f_s},
\end{equation}
where $f$ is the frequency of the detector frame, the prefactor 0.7 accounts for the spin distribution where we have taken the assumption that $70\%$ of SMBHs with initial spin $a=0.99$, $\rho_c=3H_0^2/(8\pi G)$ is the critical energy density of the Universe, the Hubble constant $H_0=100~h_{100}~\text{km}~\text{s}^{-1}~\text{Mpc}^{-1}$ with $h_{100}=0.678$ and $d E_s/d f_s$ is the energy spectrum of the radiation in the source frame. We assume $d \dot{n}/(d M_b )=(1/t_0)(dn/dM_b)$ and the derivative of the lookback time with respect to redshift is
\begin{equation}
dt/dz=\frac{1}{H_0 (1+z)\sqrt{\Delta(z)}},    
\end{equation}
where $\Delta(z)=\Omega_M(1+z)^3+\Omega_\Lambda$, $\Omega_M$ is the matter density of the universe and $\Omega_\Lambda$ is the dark energy density.

For discrete frequencies GW spectra, we have
\begin{equation}
	\frac{d E_s}{d f_s} \approx E_{\mathrm{GW}} \delta\left(f(1+z)-f_s\right),
\end{equation}
where $E_{\text{GW}}$ is the radiation energy of the GWs during the signal duration $\Delta t$ with $\Delta t=\text{min}(\tau_{\text{GW}},t_0-t_{\star},t_S-t_{\star})$. For the sake of the finite frequency resolution and signal duration of the detector, the Dirac delta function is spread out over a size $\sim$ max$[1/(\Delta t (1+z)),1/T_{\text{obs}}]$~\cite{Brito:2017zvb} where we take $T_{\text{obs}}\approx 15~\text{yr}$ for PTA experiments. And the radiation energy $E_{\text{GW}}$ can be derived from Eq.~(\ref{E}),
\begin{equation}
E_{\text{GW}}=(N_f(t_{\star}+\Delta t)
-N_f(t_{\star}))\Delta \omega.  
\end{equation}

\begin{figure}[!htp]
	\begin{center}
                \includegraphics[width=0.8\linewidth]{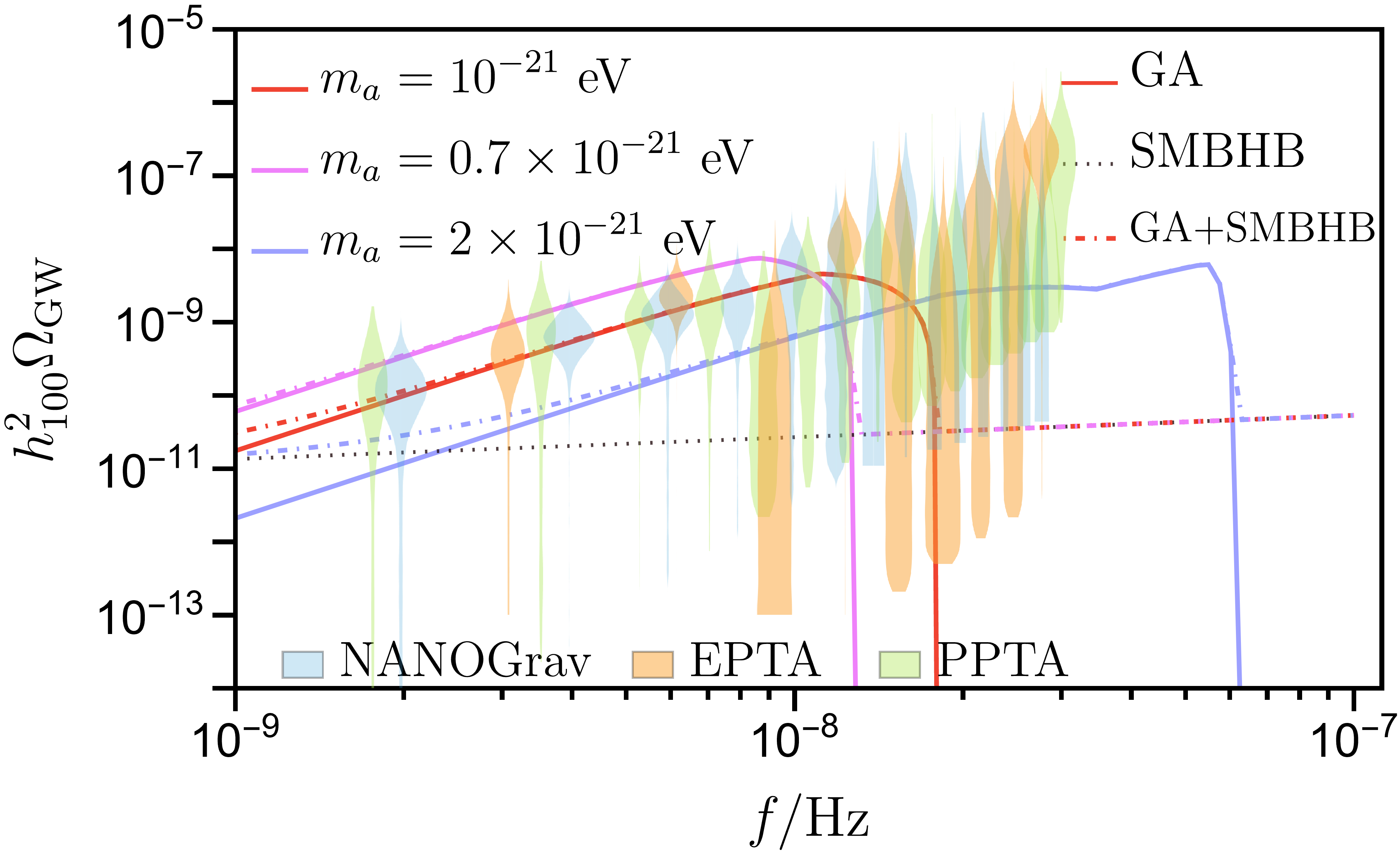}
			\caption{The total SGWB spectra from the axion transition processes for different axion masses with (dash-dotted lines) and without (solid lines) considering the contributions of SMBHBs. The red, pink, and violet lines depict the GW spectra for the axion masses of $10^{-21}~\text{eV}$, $0.7\times 10^{-21}~\text{eV}$, and  $2\times10^{-21}~\text{eV}$, respectively. The black dotted line shows the SGWB spectrum from astrophysical inspiraling SMBHBs. The colored violin plots represent the marginalized posteriors of GW energy density from the data analyses of different PTA experiments. 
   }\label{fig:gw}
		\end{center}
\end{figure}

In Fig.~\ref{fig:final},
we present the predicted GW strength originating from axion transition processes  $|644\rangle\rightarrow |544\rangle$ and $|533\rangle\rightarrow |433\rangle$ in the parameter space of the axion mass and the SGWB frequency. We consider the axion mass range $10^{-21}\sim 10^{-19}~\text{eV}(10^{-22}\sim 10^{-20}~\text{eV})$ and the frequency range $10^{-9}\sim 10^{-7}~\text{Hz}$ for the transition process $|644\rangle\rightarrow |544\rangle(|533\rangle\rightarrow |433\rangle)$. The GW signals of the mass range around $10^{-21}\sim 10^{-20}~\text{eV}(10^{-22}\sim 10^{-21}~\text{eV})$ is stronger than other mass ranges for the transition process $|644\rangle\rightarrow |544\rangle(|533\rangle\rightarrow |433\rangle)$ which can be seen from Fig.~\ref{fig:final}, and thus they may fit PTA data better. For the lighter axion mass region around $10^{-21}~\text{eV}(10^{-22}~\text{eV})$, since the energy level $|644\rangle\rightarrow |544\rangle(|533\rangle\rightarrow |433\rangle)$ we considered in the transition process dominates the GW emission at $1.38\lesssim\alpha\lesssim1.79(0.90\lesssim\alpha\lesssim1.34)$, only the very large mass BHs could satisfy this requirement. But the superradiant rate is slower for larger BH mass as we can see from Fig.~\ref{fig:sr}, thus these BHs can not produce sufficient axions through the superradiant process to generate observable SGWB strength. For the heavier axion mass region above $10^{-20}~\text{eV}(10^{-21}~\text{eV})$, the corresponding BH masses which satisfy the requirement are small. Since the axion occupation number is proportional to the square of the BH mass as Eq.~\eqref{Nm}, these small mass BHs can not produce enough axions too. Thus the energy-level transitions yield merely feeble GW signals for heavier axion masses.  And there seems exists a horizontal discontinuity around $m_a\sim10^{-20.9}~\text{eV}(10^{-21.9}~\text{eV})$ for the transition process $|644\rangle\rightarrow |544\rangle(|533\rangle\rightarrow |433\rangle)$, this is attributed to the rapid decrease in the superradiant rate on the right side of the peak as shown in Fig.~\ref{fig:sr}.  

 Unlike the case of the axion annihilation signal where the GW frequency only depends on the axion mass, the SGWB from energy-level transition is 
not monochromatic since the astrophysical SMBHs could have a wide mass range.
Moreover, there may exist other transition levels in the axion cloud. 
Although the corresponding GW strength is small compared to the case considered in this work, they can provide GW signals contributing to SGWB spectra.

\begin{figure}[!htp]
	\begin{center}
                \includegraphics[width=0.7\linewidth]{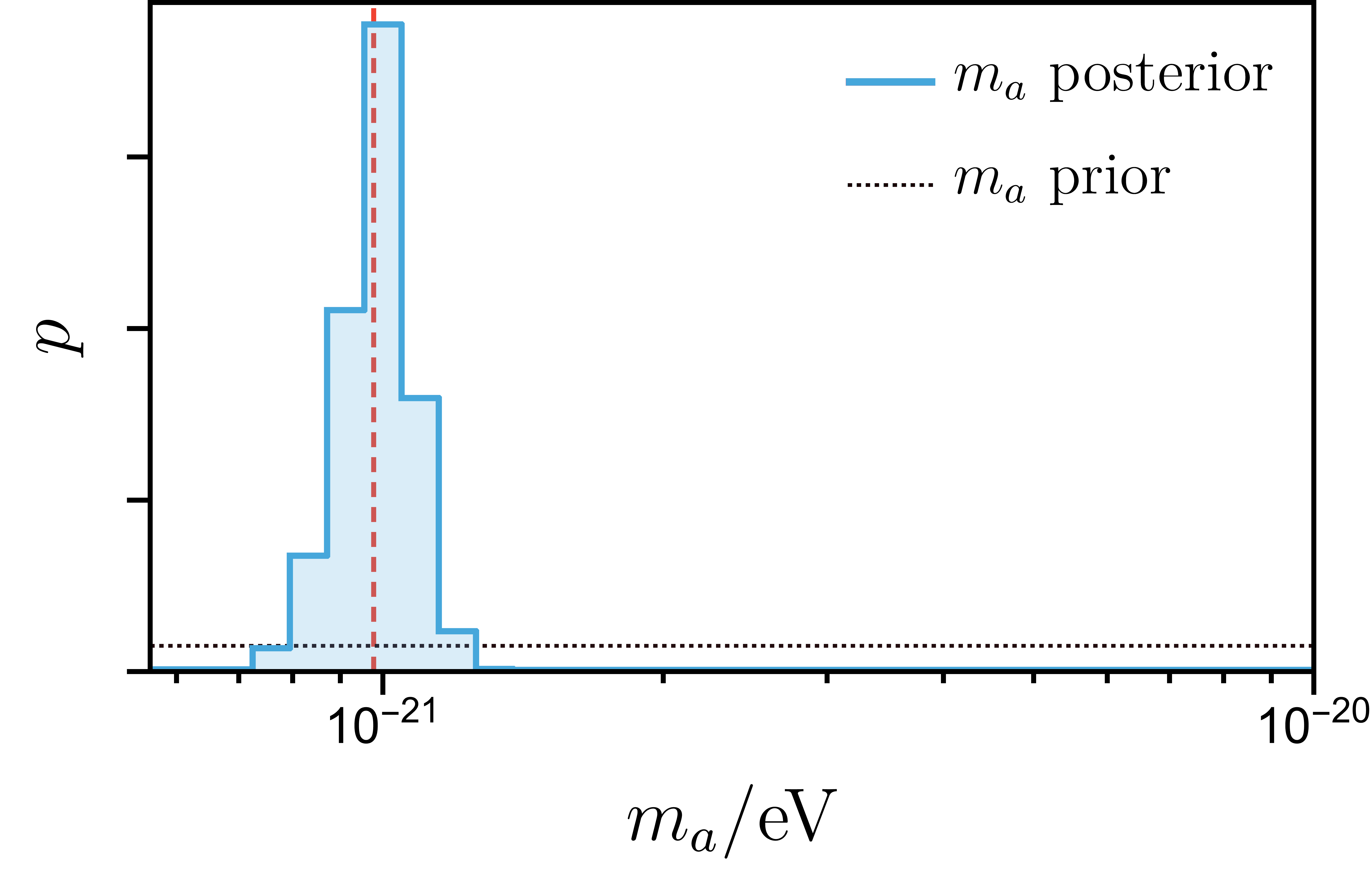}
			\caption{The posterior distribution $p$ for the axion mass $m_a$ of GA and SMBHBs combined model from NANOGrav 15-year data. 
            The posterior is averaged within the different log-uniform axion mass intervals. 
            The black dotted line depicts the log-uniform prior distribution of the axion mass. And the vertical red dashed line represents the axion mass corresponding to the maximum posterior.
   }\label{fig:posd}
		\end{center}
\end{figure}

To be more clear to see that the axion mass around $10^{-21}$~eV is favored by PTA observation, 
we show the total SGWB spectra of the two transition processes ($|533\rangle\rightarrow |433\rangle$ and $|644\rangle\rightarrow |544\rangle$) for different benchmark axion masses in Fig.~\ref{fig:gw}. 
The red, pink, and violet solid lines depict the GW spectra from axion transitions between different levels in the GA. 
The black dotted line plots the SGWB spectrum produced by the astrophysical population of inspiraling SMBHBs whose parameters are fixed at the central values of the predicted 2D-Gaussian distribution (see  Ref.~\cite{NANOGrav:2023hvm} for more details).
The colored curves show that there exists a peak frequency of SGWB and the peak frequency becomes larger when the axion mass increases. 
The SGWB strength and frequency have a relation $\Omega_{\text{GW}}\propto f^{2}$ when $f<f_{\text{peak}}$ as shown in Fig.~\ref{fig:gw}. And through the released results of PTA data, we find that the favored axion mass range is approximately around $1\times 10^{-21}~\text{eV}$. 
The Bayes factor of the combined SGWB spectra from axion transitions and SMBHBs for NANOGrav 15-year data is around \textcolor{blue}{7.1} with respect to the spectra of SMBHBs alone, obtained by \texttt{ENTERPRISE}~\cite{ep} and \texttt{ENTERPRISE}\underline{~~}\texttt{EXTENSIONS}~\cite{ee} via the wrapper \texttt{PTArcade}~\cite{Mitridate:2023oar}. 
 This indicates the axion transitions in combination with SMBHBs are more favored by NANOGrav 15-year data than SMBHBs alone. 
The posterior distribution for the axion mass $m_a$ of GA and SMBHBs combined model inferred from NANOGrav 15-year data with a log uniform prior is shown in Fig.~\ref{fig:posd}. More than $95\%$ of the overall posterior is contained in the axion mass region $10^{-21.2}\sim10^{-20.7}~\text{eV}$, indicating the NANOGrav 15-year data favors the existence of a light scalar within this mass range. 
Besides the data released by the NANOGrav, PPTA, and EPTA, the CPTA results also favor the same mass range when considering the SGWB originating from the GAs. 
It should be noted that there are other BH mass distribution functions with different power laws and normalization factors that may lead to different results. And for the axion mass  $m_a \sim 10^{-21}~\text{eV}$ favored by PTA experiments concerning the positive power index BH mass distribution model we have used, the SGWB strength is about one and a half order smaller if we use the negative power index BH mass distribution model. Thus the positive power index BH mass distribution model is more in line with current PTA data. We demonstrate the detailed discussions of other BH spin and mass distribution models in Appendix.\ref{app}.

\section{Conclusion}\label{conc}
We have studied the SGWB from axion transitions between different energy levels composed of the cosmic populated BHs and their surrounding axion clouds
formed through the superradiant process. 
Considering the error bars of the current observation at NANOGrav, EPTA, PPTA, and CPTA data, we found that this type of SGWB can naturally reproduce the observed GW spectra
if the axion mass is around $1\times10^{-21}$~eV for the positive power index BH number density distribution function with the assumption that $70\%$ of SMBHs initial spin $a=0.99$.
And for the other representative BH number density distribution function with a negative power index, the axion mass range favored by PTA data is nearly unchanged. 
The Bayesian analysis shows that the GW radiated from the transition of the GA can also explain the GW energy density posterior distributions. 
The combination of SGWB spectra generated by the GA transition and the SMBHBs is more favored for the SGWB from the SMBHBs.

The assumptions we made in this work are the spin and mass distribution of SMBHs and the existence of ultralight aixon particles around $10^{-21}$ eV, which is around the mass range of the the well-motivated fuzzy DM. Compared to the well-known interpretations like cosmic string, domain wall, and phase transition~\cite{NANOGrav:2023hvm, EPTA:2023xxk} , it is less restricted by cosmological constraints and fine-tuning problems since it is formed after the Big Bang nucleosynthesis period. Our work also shows that the PTA experiments can be a novel and realistic approach to probing the superradiance or Penrose process, 
the isolated BHs, the ultralight axion particles or the fuzzy DM. 
As the leading-order term of the overlap reduction function, the Hellings-Downs curve represents the isotropic information, while the anisotropy analysis based on the higher-order terms of the overlap reduction function at future PTA experiments and its deeper implication for the sources properties (such as the detailed distribution information of BHs) are left for our future study~\cite{Li:2021iva}.

\acknowledgments
This work was supported by the National Natural Science Foundation of China (NNSFC) under Grant No.12205387, No.12475111, and Guangdong Major Project of Basic and Applied Basic Research (Grant No.2019B030302001). 

\appendix
\section{Mass distribution of isolated black holes}\label{app}
The distribution of isolated BHs is essential in our calculations. We consider two models of mass distribution as in Refs.~\cite{Brito:2017zvb} and \cite{Babak:2017tow},

\begin{figure}[!htb]
	\begin{center}
			\includegraphics[width=0.45\linewidth]{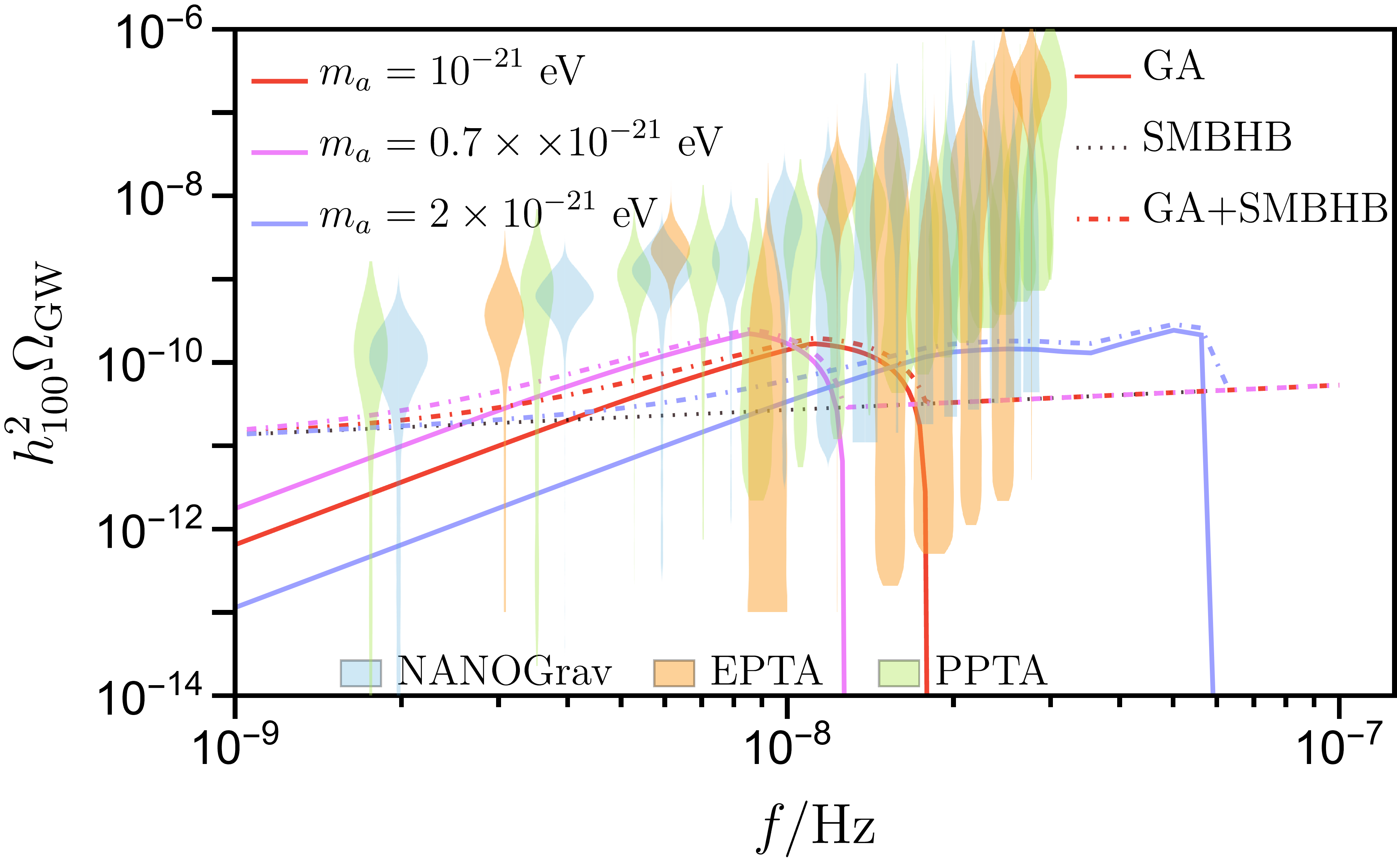}
            \includegraphics[width=0.45\linewidth]{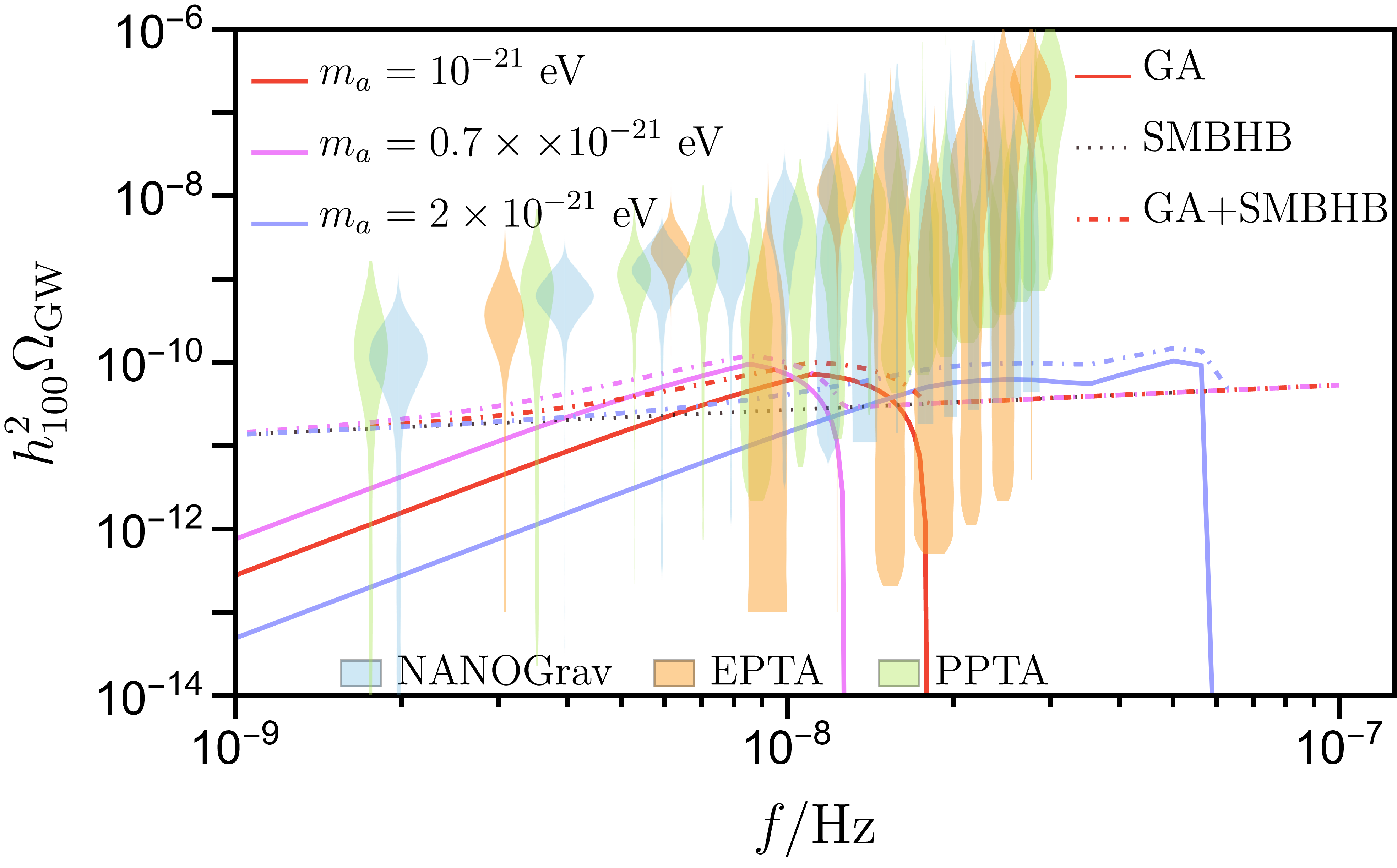}
			\caption{The total SGWB spectra from the axion transition processes for the PTA favored axion mass range with (dash-dotted lines) and without (solid lines) the SMBHB signal. Left panel: $70\%$ of SMBHs with initial spin $a=0.99$; right panel: $30\%$ of SMBHs with initial spin $a=0.99$. Here we use the negative power index BH mass distribution function, i.e.  Eq.~(\ref{distr2}). The red, pink, and violet lines depict the GW spectra for the axion masses of $10^{-21}~\text{eV}$, $0.7\times 10^{-21}~\text{eV}$, and  $2\times 10^{-21}~\text{eV}$, respectively. 
            The colored violin plots represent the marginalized posteriors of GW energy density from the data analyses of different PTA experiments. The black dotted line shows the SGWB spectrum from inspiraling SMBHBs.
   }\label{fig:gwl} 
		\end{center}
\end{figure}

(1) A positive power index analytic function
\begin{equation}\label{distr1}
 \frac{d n}{d \log_{10}M_b }=0.002\left(\frac{M_b}{3\times 10^6~M_{\odot}}\right)^{0.3}\text{Mpc}^{-3},
 \end{equation}
 where 
 $M_{\odot}$ is solar mass. This formula is used for $10^4~M_{\odot}<M_b<10^7~M_{\odot}$ and $z<3$, while for $M_b>10^{7}~M_{\odot}$, we reduce the distribution 100 times for a conservative model compared to Eq.~(\ref{distr1}).
\\

(2) A negative power index analytic function
\begin{equation}\label{distr2}
 \frac{d n}{d \log_{10}M_b}=0.005\left(\frac{M_b}{3\times 10^6~ M_{\odot}}\right)^{-0.3}\text{Mpc}^{-3},
 \end{equation}
This formula is used for $10^4~M_{\odot}<M_b<10^7~M_{\odot}$ and $z<3$, while for $M_b>10^{7}~M_{\odot}$, we reduce the distribution 10 times compared to Eq.~(\ref{distr2}).

We have used the mass function in Eq.~(\ref{distr1}) to estimate the SGWB in Sec.~\ref{sec3}. And here we give the results of the other mass distribution model as in Eq.~(\ref{distr2}). As a more realistic comparison, we also give the results that only $30\%$ of SMBHs have extreme initial spin in this model. The SGWB energy density of different axion masses and frequencies with different BH spin distributions concerning the negative power index mass distribution function is shown in Fig.~\ref{fig:gwl}. The red, pink, and violet lines depict the  GW spectra for the axion masses of $10^{-21}~\text{eV}$, $0.7\times 10^{-21}~\text{eV}$, and  $2\times10^{-21}~\text{eV}$, respectively. For the axion mass around $1\times10^{-21}~\text{eV}$, the SGWB strength of the negative power index distribution model is nearly one and a half order smaller than the positive index distribution model. We find that the axion mass range favored by PTA data is roughly unchanged for different BH mass distribution models.
And the SGWB energy density can be ignored below the axion mass $10^{-22}$~eV for both SMBH distribution models. 

\bibliographystyle{JHEP}
\bibliography{ren.bib}

\end{document}